\documentclass[12pt]{extarticle}

\usepackage{geometry}
\usepackage{setspace}
\usepackage{arxiv}
\usepackage{lmodern}  % For improved font compatibility
\usepackage[T1]{fontenc}
\usepackage[utf8]{inputenc}

\usepackage{graphicx}
\usepackage{multirow}
\usepackage{amsmath,amssymb,amsfonts}
\usepackage{amsthm}

\usepackage[title]{appendix}
\usepackage{xcolor}
\usepackage{textcomp}
\usepackage{manyfoot}
\usepackage{booktabs}
\usepackage{algorithm}
\usepackage{algorithmicx}
\usepackage{algpseudocode}
\usepackage{listings}

\usepackage{hyperref}       % hyperlinks
\usepackage{url}            % simple URL typesetting
\usepackage{booktabs}       % professional-quality tables
\usepackage{amsfonts}       % blackboard math symbols
\usepackage{amsmath,amssymb}
\usepackage{nicefrac}       % compact symbols for 1/2, etc.
\usepackage{microtype}      % microtypography
\usepackage{graphicx}
\usepackage[numbers]{natbib}

\usepackage{doi}

\title{Nonlinear Dissipative Forces in Celestial Motion Using the Method of Multiple Scales}

\author{Raju S. Khatiwada\textsuperscript{1}\thanks{Corresponding author. \texttt{r.khatiwada@goldengate.edu.np}} , 
C. Ortiz\textsuperscript{2}, 
Basanta R. Giri\textsuperscript{1} \\[1em]
\textsuperscript{1} GoldenGate International College, Tribhuvan University, Kathmandu, 44600, Nepal. \\[0.5em]
\textsuperscript{2}Unidad Académica de Física, Universidad Autónoma de Zacatecas, \\ Calzada Solidaridad esquina con Paseo a la Bufa S/N, Zacatecas, 98060, México.
}

\renewcommand{\shorttitle}{Nonlinear Dissipative Forces in Celestial Motion Using Method of Multiple Scales}

\hypersetup{
    pdftitle={Nonlinear Dissipative Forces in Celestial Motion Using the Method of Multiple Scales},
    pdfsubject={Astrophysics},
    pdfauthor={Raju S. Khatiwada, C. Ortiz, Basanta R. Giri},
    pdfkeywords={Gravitational Friction, Precession of Perihelion of Mercury, General Relativity},
}

\begin{document}

\maketitle

\begin{abstract}
This paper investigates the influence of nonlinear dissipative forces, specifically Gravitational Friction (GF), on the precession of celestial bodies within the framework of general relativity. We derive a modified line element by introducing a density-dependent term to model interactions between planetary bodies and the low-density interplanetary medium, providing a covariant description of dissipative forces in planetary motion. The resulting metric modification leads to corrections in the perihelion precession of Mercury, also reproducing the classical relativistic predictions. Utilizing the method of multiple scales, we analyze perturbative effects induced by GF. Using this model, we successfully constrain the medium density near Mercury to approximately $\rho_0 \approx 1.12 \times 10^{-10} \, \text{kg/m}^3$. These findings offer a new approach for incorporating dissipative mechanisms into general relativity, with potential applications in other astrophysical systems.
\end{abstract}

% Keywords

\section{Introduction}\label{sec1}

Dynamical friction is the decelerative force exerted on a massive object as it moves through a medium, producing a gravitational wake that pulls back on the object. This force is linearly proportional to the object's velocity and mass, as described by Chandrasekhar’s dynamical friction formula~\cite{chandrasekhar1943}. While effective in predicting orbital decay in many astrophysical contexts, dynamical friction assumes a homogeneous and isotropic medium and neglects the more complex interactions that can arise in real celestial environments~\cite{just1990}. These assumptions limit its applicability, particularly in regions where the medium is highly anisotropic or involves nonlinear dissipative forces.

In astrophysics, photon-medium interactions are well-documented and play a crucial role in understanding phenomena like redshift. However, the motion of celestial objects, such as planets, through a medium with varying orbital parameters remains comparatively less explored. For instance, planets migrating through protoplanetary disks experience dynamical friction, where gravitational interactions with the gas and dust in the disk create a drag force, leading to orbital decay or inward migration~\cite{1997Icar..126..261W, 1980ApJ...241..425G}. Similarly, gravitational drag occurs when planets move through low-density interstellar or interplanetary mediums, inducing a decelerative force over long periods~\cite{1999ApJ...513..252O, 1978ppim.book.....S}. 

Poynting-Robertson drag, for example, causes dust particles in planetary systems to spiral inward due to stellar radiation~\cite{1979Icar...40....1B}. Tidal forces between planets and nearby massive bodies, such as stars or moons, lead to tidal dissipation within planetary interiors, modifying both rotational and orbital parameters over time~\cite{1966Icar....5..375G, 1978Icar...36..245P}.

In the framework of relativistic planetary motion, especially in highly eccentric orbits or near relativistic regimes, density gradients or perturbations in the interstellar or circumstellar medium may induce additional drag forces that vary nonlinearly with velocity and spatial position~\cite{1999ApJ...513..252O,spitzer1978}. These nonlinear dissipative forces could alter orbital parameters and impact phenomena like the precession of planetary orbits, with corrections beyond classical relativistic predictions~\cite{will2014}.

Einstein's General Theory of Relativity (GR) has historically presented challenges in verification due to the theory's complexity and the difficulty of performing direct experimental tests. One of the most well-known confirmations of GR came through gravitational lensing observations. In 1919, Sir Arthur Eddington and astronomer Frank Watson Dyson~\cite{1920RSPTA.220..291D} observed a solar eclipse, measuring the apparent shift in star positions near the Sun due to its gravitational field.

Understanding celestial mechanics also requires considering the interaction of celestial objects with the interstellar and interplanetary media, as these interactions alter the objects' orbital parameters. For instance, Mercury's orbit precesses due to perturbations from other solar system bodies, the Sun's oblateness, and non-Newtonian gravitational effects~\cite{1978bah..book.....R}.

In a previous work by Ortiz-Khatiwada~\cite{Ortiz2023}, it was shown that applying d’Alembert’s principle of virtual work provides a formal method for establishing energy dissipation due to gravitational fields in a low-density medium when non-holonomic constraints are present. This phenomenon is termed Gravitational Friction (GF). 

In this work, we extend the concept of gravitational friction  as a dissipative force within the framework of general relativity. This is accomplished by deriving a modified line element that incorporates dissipative interactions with the low-density interplanetary medium through a density-dependent term representing the effects of GF. In Section \ref{sec2} we introduce the concept of gravitational friction, its relation with the redshift parameter and we use this relation to eventually derive a line element that includes the gravitational interaction as well the dissipative effects due to GF. In Section \ref{sec3}, we establish the dynamics of the proposed system by systematically applying the method of multiple scales. This approach allows for the inclusion of first-order corrections, enabling a detailed analysis of their impact on perihelion precession. Finally, in Section \ref{sec4}, we present numerical results, validate the model against observational data for Mercury, and constrain the medium density near Mercury.
\section{Gravitational Friction and Redshift}\label{sec2}
The gravitational friction mechanism~\cite{Ortiz2023} describes energy dissipation that occurs when an object moves through a surrounding low-density medium. Based on d’Alembert’s principle, it involves the virtual work done by gravitational forces during the object’s motion. The energy lost through these interactions can be analyzed by treating the system as if it were in equilibrium, with inertial forces accounting for the dynamic behavior. This approach provides a framework to study non-conservative forces and their effects on the motion of objects in various astrophysical and cosmological contexts.

The work done by the particle on the medium, as a result of the gravitational friction mechanism, can be calculated by integrating the force contributions along its path, yielding the expression:

\begin{equation} \label{wo}
W = -\frac{1}{3} \frac{hf}{c^2} G \pi \rho_0 R^2. 
\end{equation} 
Here,  $f$  represents the photon’s frequency, $ \rho_0$  is the average density of the medium along the particle’s path, and  $R$  is the distance traveled by it. The energy lost by the particle, caused by interactions with the medium due to gravitational friction, is transferred to the medium as energy.

% This energy loss represents an actual decrease in the photon’s energy, which could be misinterpreted as a shift resulting from a change in the frame of reference.

The energy lost by the particle due to the work done on the medium, as described by the gravitational friction mechanism, can be directly related to the redshift using the generalized redshift formula within an energy-based framework \cite{redshft}. In this generalization, the redshift  $z$  is defined as the fractional change in the square root energy of the particle.
\begin{equation} \label{1+z}
 1+z\equiv \frac{ \mathcal{E}_\mathrm{e}^{1/2}}{\mathcal{E}_{\mathrm{o}}^{1/2}},
\end{equation}
where  $\mathcal{E}_e={h^2 f}^2/2p^2$ represents the specific energy emitted in the ground state in terms of the Planck's constant $h$, the frequency of the particle $f$, and the particle's momentum $p$. While the observed specific energy in the  considered gravitational potential and the energy loss due to the gravitational friction work (\ref{wo}) is given by:
\begin{equation}
     \mathcal{E}_o=\frac{{h^2 f}^2}{2p^2}+\frac{GM}{r}+\frac{1}{3} G \pi \rho_0 R^2,
\end{equation}
where the first term in the right-hand side of the equation arises from the kinetic energy, the second term from the gravitational potential, and the last term from the dissipative interaction with the medium.
The redshift relation for the described system is given by:
\begin{equation}\label{6}
    1+z= \frac{1}{\sqrt{1-2\frac{GM}{rc^2}-\frac{2 }{3{c^2}}\pi G\rho_0 R^2}},
\end{equation}
for a particle traveling in a the radial coordinate $r=R$.

Given Weyl's general redshift in terms of the line element, 
\begin{equation}
    1+z=\frac{ds_o}{ds_e},
\end{equation}
$ds_e$ and $ds_r$ are the worldline elements relative to the emitter and the observer.  The redshift relation can be recast in terms of the temporal component of the metric tensor.

\begin{equation}\label{timedilation}
    1+z= \sqrt{\frac{g_{tt} (\text{o})}{g_{tt} (\text{e})}}.
\end{equation}
This formula emerges because  $g_{tt}$  determines the rate of proper time passage for static observers in the spacetime. Therefore, we can relate the line element with the gravitational field and the gravitational friction term by:  
\begin{equation}\label{7}
    g_{tt}(e)=1-2\frac{GM}{rc^2}-\frac{2 }{3{c^2}}\pi G\rho_0 R^2.
\end{equation}

To incorporate dissipativity due to GF into General Relativity, we must work within a covariant framework consistent with the principles of General Relativity. While the Tolman-Komar Energy relation provides a useful tool for understanding the total energy of a gravitational system, it may not directly address the inclusion of dissipative effects.

%In a cosmological scenario, modifications to relevant equations, such as the Friedmann equation, are necessary to account for the dissipative effects due to GF. We need to select the corresponding line element that includes such a characteristic. One approach is to introduce an additional term on the left-hand side of the equation, as suggested by Ortiz \cite{Ortiz1}. The expression for GF from Ortiz et al. \cite{Ortiz2023} is to be used to get the modified Friedmann equation as:
Considering the modified temporal metric tensor (\ref{7}), and by assuming spherical symmetry, the line element can be cast as
\begin{equation}   \label{8}
ds^2 = -\left(1 - \frac{2GM}{c^2r}-\frac{2 }{3{c^2}}\pi G\rho_0 R^2\right)c^2dt^2 + \frac{dr^2}{1 - \frac{2GM}{c^2r}-\frac{2 }{3{c^2}}\pi G\rho_0 R^2} + r^2d\Omega ^{2},
\end{equation}
here, $d\Omega ^{2}$ is the standard metric on the 2-sphere. The line element contains the information due to the gravitational field and the dissipative term arising from the motion of the object through a medium, i.e., gravitational friction. From the explicit line element, it is possible to derive the dynamics of a particle. 

In the present set-up, where we analyze a celestial body of mass $\mu$ which revolves in its orbit with some effective observational velocity, we can relate the distance traversed $R$ with the radial distance from the source $r$. 
%Now, we modify all other equations based on this contribution.

\section{Dynamics}\label{sec3}
In this section, we construct the dynamical framework for a particle influenced by a medium with density-dependent effects . Starting from a Lagrangian formulation, we derive the associated Hamiltonian, incorporating corrections that account for the dissipative interactions due to the density of the medium. We use the method of multiple scales to solve the resulting equations, systematically integrating first-order corrections into the equations of motion. 

Departing from the modified line element (\ref{8}), the Lagrangian for a mass particle $\mu$  can be cast as,

\begin{equation}
\begin{aligned}
L = & \frac{\mu}{2}\biggl[ \left( 1 - \frac{R_s}{r} - \frac{2G\pi\rho_0 r^2}{3c^2} \right) c^2 \dot{t}^2 \\
& - \left(1 - \frac{R_s}{r} - \frac{2G\pi\rho_0 r^2}{3c^2}\right)^{-1} \dot{r}^2 \\
& - r^2\left( \dot{\theta}^2 + \sin^2\theta \dot{\phi}^2 \right) \biggr],
\end{aligned}
\label{corrected_lagrangian}
\end{equation}
where $\rm R_s=\frac{2GM}{c^2}$ is the Schwarzschild radius of the Sun,  $\rm m_s = M $ is the Sun's mass, and the traversed distance $R \propto r$.

The corresponding momenta are obtained in the usual way
\begin{eqnarray}
\rm \Pi_{t} &=& \rm \frac{\partial {\cal L}}{\partial \dot{t}} = \mu\left( 1 - \frac{R_s}{r} - \frac{2G\pi\rho_0 r^2}{3c^2} \right) c^2 \dot{t}, \label{pit_corrected} \\
\rm \Pi_{r} &=& \rm \frac{\partial {\cal L}}{\partial \dot{r}} = -\mu\left( 1 - \frac{R_s}{r} - \frac{2G\pi\rho_0 r^2}{3c^2} \right)^{-1} \dot{r}, \label{pir_corrected} \\
\rm \Pi_{\theta} &=& \rm \frac{\partial {\cal L}}{\partial \dot{\theta}} = -\mu r^{2} \dot{\theta}, \label{pitheta_corrected} \\
\rm \Pi_{\phi} &=& \rm \frac{\partial {\cal L}}{\partial \dot{\phi}} = -\mu r^2 \sin^2 \theta \dot{\phi}, \label{piphi_corrected}
\end{eqnarray}

so, we build the Hamiltonian as

\begin{equation}
\rm H=\Pi_{q^i} \, \dot q^i-{\cal L}.
\end{equation}

For simplicity, we consider the plane $\rm \theta=\frac{\pi}{2}$, then, $\rm \dot \theta= \ddot \theta=0.$ Finally, the Hamiltonian function becomes

\begin{equation}
\begin{aligned}
H = & \frac{1}{2\mu} \biggl[ \frac{1}{c^2} \left( 1 - \frac{R_s}{r} - \frac{2G\pi\rho_0 r^2}{3c^2} \right)^{-1} \Pi_{t}^{2} 
& - \left( 1 - \frac{R_s}{r} - \frac{2G\pi\rho_0 r^2}{3c^2} \right) \Pi_{r}^{2} - \frac{\Pi_{\phi}^{2}}{r^{2}} \biggr].
\end{aligned}
\label{hamiltonian_corrected}
\end{equation}

The dynamics of the system are given by

\begin{eqnarray}
\dot{r} &=& -\frac{\Pi_r}{\mu} \left(1 - \frac{R_s}{r} - \frac{2G\pi\rho_0 r^2}{3c^2}\right), \label{dr_corrected} \\
\dot{\Pi}_r &=& -\frac{1}{\mu} \biggl[ \frac{\frac{R_s}{r^2} - \frac{2G\pi r^2}{c^2}}{c^2 \left(1 - \frac{R_s}{r} - \frac{2G\pi\rho_0 r^2}{3c^2}\right)^2} \Pi_t^2 \nonumber \\
&& + \left(\frac{R_s}{r^2} - \frac{2G\pi r^2}{c^2}\right) \Pi_r^2 - \frac{\Pi_\phi^2}{r^3} \biggr], \label{dpir_corrected} \\
\dot{t} &=& \frac{1}{\mu c^2} \left(1 - \frac{R_s}{r} - \frac{2G\pi \rho_0 r^2}{3c^2}\right)^{-1} \Pi_t, \label{dt_corrected} \\
\dot{\Pi}_t &=& 0, \label{dpit_corrected} \\
\dot{\phi} &=& -\frac{\Pi_\phi}{\mu r^2}, \label{dphi_corrected}\\
\dot{\Pi}_\phi &=& 0. \label{dpiphi_corrected}
\end{eqnarray}

From equations (\ref{dpit_corrected}) and (\ref{dpiphi_corrected}), we obtain:
\begin{equation} \label{22}
\Pi_t = \alpha, \qquad \Pi_\phi = \ell,
\end{equation}
where $\alpha$ and $\ell$ are the constants of motion.

Thus, if the total energy is $E$, then from equation (\ref{hamiltonian_corrected}), it follows that:
\begin{equation} \label{energy}
2\mu E = \left[ \frac{\alpha^2}{c^2}\left( 1 - \frac{R_s}{r} - k r^2 \right)^{-1} - \left( 1 - \frac{R_s}{r} - k r^2 \right) \Pi_r^2 - \frac{\ell^2}{r^2} \right],
\end{equation}
where $k = \frac{2 G \pi \rho_0}{3 c^2}$.

The term for $2\mu E$ is:
\begin{equation}
2\mu E = \left\{
\begin{array}{ll}
\mu^2 c^2 & \text{for particles} \\
0 & \text{for photons}
\end{array}
\right\}.
\label{energia_corrected}
\end{equation}

Substituting the value of $\Pi_{r}$ from equation (\ref{pir_corrected}) in (\ref{energy}), we have:

\begin{equation}
\begin{aligned}
\mu^2 c^2 = & \left[ \frac{\alpha^2}{c^2}\left(1-\frac{R_s}{r}-kr^2\right)^{-1} \right. \\
& -\mu^2\dot{r}^2\left(1-\frac{R_s}{r}-kr^2\right)^{-1} -\frac{l^2}{r^{2}}\biggr].
\end{aligned}
\end{equation}

Rearranging gives,
\begin{equation} \label{26}
\rm \dot{r}^{2}=\rm \frac{1}{\mu^2}\left[ \frac{\alpha^2}{c^2}-\frac{l^2}{r^2}\left(  1-\frac{R_s}{r}-kr^2\right)-\mu^2 c^2\left(  1-\frac{R_s}{r}-kr^2\right) \right].
\end{equation}
To keep things simple, we make the change of variable from r to $\frac{1}{u}$

\begin{equation*}\label{rdot}
 u=\frac{1}{r},\qquad \to \qquad  \dot  r=\frac{dr}{dt}=\frac{d}{dt}\left(  \frac{1}{u}\right)  =
-\frac{1}{u^2}\frac{du}{d\phi}\, \dot \phi .
\end{equation*}
With this change of variables and after substituting the value of $\dot \phi$ from equation (\ref{dphi_corrected}) and (\ref{22}) in (\ref{26}), we get,

\begin{equation}  \label{27}
\begin{aligned}
\left( \frac{du}{d\phi} \right)^2 &= \frac{1}{\ell^2} \biggl[ \frac{\alpha^2}{c^2} - \ell^2 u^2 \left( 1 - R_s u - \frac{k}{u^2} \right) \\
&\quad - \mu^2 c^2 \left( 1 - R_s u - \frac{k}{u^2} \right) \biggr].
\end{aligned}
\end{equation}

Taking the differential gives,

\begin{equation}  \label{master1}
    \frac{d^2 u}{d\phi^2} + u = \frac{3m_sG}{c^2}u^2 + \frac{\mu^2 m_s G}{l^2} - \frac{\mu^2c^2k}{l^2}\frac{1}{u^3},
\end{equation}

we perform the change of variable
\begin{equation*}
u=\frac{\mu^2m_sG}{l^2}v=m v,
\end{equation*}
where $m =\frac{\mu^2m_sG}{l^2} $

equation (\ref{master1}) becomes
\begin{equation}
   m\frac{d^2v}{d\phi^2}+mv -m -\frac{3{m_s}G}{c^2}m^2 v^2 +\frac{\mu^2 c^2k m }{l^2 m^4 v^3}=0, \label{32}
\end{equation}

\begin{equation}
    \frac{d^2v}{d\phi^2}+v -1-\epsilon v^2+\frac{\delta}{v^3}=0 ,\label{finaleq}
\end{equation}
where $\epsilon$, $\delta$ and are given by equation  (\ref{31})

\begin{equation}
\left\{
\begin{array}{l}
\epsilon = \frac{3\mu^2{m_s}^2G^2}{l^2c^2} \\[8pt]
\delta = \frac{\mu^2 c^2 }{l^2 }\frac{k}{m^4} = \frac{2\pi l^6 \rho_0}{3\mu^6 m_s^4 G^3}
\end{array}
\right\}.
\label{31}
\end{equation}

Equation (\ref{finaleq}) is our main equation. We solve it by employing the technique of multiple scales. Let,
\begin{equation}
    v= \sum_{i,j=0}^{\infty}v_{ij}\epsilon^i\delta^j,
\end{equation}

\begin{equation}
    \Phi_{i,j}=\epsilon^i\delta^j\phi.
\end{equation}
So by the chain rule, 
\begin{equation}
    \frac{d}{d\phi} = \sum_{i,j=0}^{\infty} \epsilon^i \delta^j \frac{\partial}{\partial \Phi_{i,j}}.
\end{equation}

Then equation (\ref{finaleq}) becomes,

\begin{multline}
\begin{aligned}
& \left(\frac{\partial}{\partial \Phi_{00}} + \epsilon \frac{\partial}{\partial \Phi_{10}} + \delta \frac{\partial}{\partial \Phi_{01}} + \dots \right)^2 \left(v_{00} + \epsilon v_{10} + \delta v_{01} + \dots \right) \\
& + \left(v_{00} + \epsilon v_{10} + \delta v_{01} + \dots \right) - 1 \\
& - \epsilon \left(v_{00} + \epsilon v_{10} + \delta v_{01} + \dots \right)^2 \\
& + \delta \left(v_{00} + \epsilon v_{10} + \delta v_{01} + \dots \right)^{-3} = 0
\end{aligned}
\end{multline}

For the first part, we find the equation of order $\epsilon^0\delta^0$ which is given as,
\begin{equation}  \label{00}
    \frac{\partial^2 v_{00}}{\partial \Phi_{00}^2} + v_{00} - 1 = 0.
\end{equation}

The solution of equation (\ref{00}) is
\begin{equation}
     v_{00}=1+Be^{i\Phi_{00}}+ \bar Be^{-i\Phi_{00}}, \label{37}
\end{equation}
where $B$ is a complex-valued function independent of $\Phi_{00}$

The equation of order $\epsilon^1\delta^0$ is

\begin{equation}
    \frac{\partial^2 v_{10}}{\partial \Phi_{00}^2} + v_{10} =  - \frac{\partial ^2 v_{00}}{\partial \Phi_{00} \partial \Phi_{10}} - \frac{\partial ^2 v_{00}}{\partial \Phi_{10} \partial \Phi_{00}}  + v_{00}^2.
\end{equation}

To solve this equation, we build on the solution of (\ref{00}). We choose $B$ and $\bar B$ in (\ref{37}) such that the resonant terms cancel out. If the resonant terms (terms proportional to  $e^{i\Phi{00}}$ and $e^{-i\Phi{00}}$) are allowed to persist, they can cause the amplitude of B to grow linearly or faster with $\Phi$, leading to unbounded behavior. Therefore, we must have,
\begin{equation}
    -2i\frac{\partial B}{\partial \Phi_{10}}+2B=0, \label{39}
\end{equation}
and its complex conjugate is
\begin{equation}
    2i\frac{\partial \Bar{B}}{\partial \Phi_{10}}+2\Bar{B}=0. \label{40}
\end{equation}
The solution to (\ref{39}) is  $B=be^{-i\beta}$ where b is the real-valued amplitude of the oscillatory part of the solution and $\beta$ is the phase or argument that gives the phase shift. They are typically determined from initial or boundary conditions.

Rearranging equation (\ref{39}) and solving gives \( b \) and \( \beta \):
\begin{equation}
\frac{\partial B}{\partial \Phi_{10}} = i B.
\end{equation}

Taking the derivative of \( B = b e^{-i \beta} \) with respect to \( \Phi_{10} \), we get:
\begin{equation}
    \frac{\partial B}{\partial \Phi_{10}} = \frac{\partial}{\partial \Phi_{10}} \left( b e^{-i \beta} \right) = -i b e^{-i \beta} \frac{\partial \beta}{\partial \Phi_{10}}.
\end{equation}

From the original equation, we have:
\begin{equation}
    \frac{\partial B}{\partial \Phi_{10}} = i B = i b e^{-i \beta}.
\end{equation}

Equating the two expressions:
\begin{equation}
    -i b e^{-i \beta} \frac{\partial \beta}{\partial \Phi_{10}} = i b e^{-i \beta}.
\end{equation}

Canceling the common terms \( b e^{-i \beta} \) from both sides, we obtain:
\begin{equation}
    \frac{\partial \beta}{\partial \Phi_{10}} = 1,
\end{equation}

\begin{equation}
    \frac{\partial b}{\partial \Phi_{10}} = 0 \implies b \ne f(\Phi_{10}),
\end{equation}

\begin{equation}
\beta = \Phi_{10} + \gamma, 
\end{equation}
where \( \gamma \) is an arbitrary phase shift independent of both \( \Phi_{00} \) and \( \Phi_{10} \).

Also, it is important to note that b is a real-valued amplitude constant with respect to $\Phi_{10}$.

The equation of order $\epsilon^0\delta^1$ is:
\begin{equation} \label{41}
    \frac{\partial^2 v_{01}}{\partial \Phi_{00}^2} + v_{01} = -\frac{\partial^2 v_{00}}{\partial \Phi_{00} \partial \Phi_{01}} -\frac{\partial^2 v_{00}}{\partial \Phi_{01} \partial \Phi_{00}}  -\frac{1}{v_{00}^3}.
\end{equation}

To know the conditions such that the resonant terms cancel out, we find the expression for $\frac{1}{{v_{00}}^3}$ first. Note that ${v_{00}}$ is given by (\ref{37}).
\\
\\
\textbf{Finding value of }$\frac{1}{v_{00}^3}$
\\
The relevant expression is:

\begin{equation}
    \frac{1}{v_{00}^3} = \left(1+Be^{i\Phi_{00}}+ \bar Be^{-i\Phi_{00}} \right)^{-3}.
\end{equation}
Now, using the substitution \( e^{i \Phi_{00}} = z \), we get:
\begin{equation}
    \frac{1}{v_{00}^3} = \left( 1 + Bz + \bar{B} z^{-1} \right)^{-3}.
\end{equation}
Using contour integration for the above, we get:
\begin{equation}
    \frac{1}{2\pi} \int_0^{2\pi} e^{-i \Phi_{00}} \left( 1 + B e^{i \Phi_{00}} + \bar{B} e^{-i \Phi_{00}} \right)^{-3} d\Phi_{00}.
\end{equation}

The denominator simplifies as:
\begin{equation}
    1 + Bz + \bar{B} z^{-1} = \frac{B z^2 + z + \bar{B}}{z}.
\end{equation}

Thus, the contour integral becomes:
\begin{equation}
    \frac{1}{2\pi i} \int \frac{z^2 dz}{\left( B z^2 + z + \bar{B} \right)^3} = \frac{1}{B^3} \frac{1}{2\pi i} \int \frac{z^2 \, dz}{(z - \alpha)^3 (z - \beta)^3}.
\end{equation}

The roots of the quadratic equation \( B z^2 + z + \bar{B} = 0 \) are denoted as \( z_1 = \alpha \) and \( z_2 = \beta \), where \( |\alpha| < 1 \) and \( |\beta| > 1 \). The pole of order 3 is at \( z = \alpha \), so we focus on this pole.

The residue at \( z = \alpha \) is given by:

\begin{equation}
    \text{Res}(f, \alpha) = \frac{1}{(n-1)!} \lim_{z \to \alpha} \frac{d^{n-1}}{dz^{n-1}} \left[ (z - \alpha)^n f(z) \right].
\end{equation}
For n= 3, it becomes:
\begin{equation}
    \text{Res}(f, \alpha) = \frac{1}{2!} \lim_{z \to \alpha}  \frac{d^2}{dz^2}\left(  \frac{z^2}{(z - \beta)^3} \right).
\end{equation}

Upon differentiating and simplifying, it yields:
\begin{equation}
    \text{Res}(f, \alpha) = \frac{B^3(1 + 2B \bar{B})}{ (1 - 4B \bar{B})^\frac{5}{2}}.
\end{equation}
 The contour integral leads to:
\begin{equation}
    \frac{1}{v_{00}^3} =\frac{1}{B^3} \frac{1}{2\pi i}. \text{Res}(f, \alpha) = \frac{1}{2\pi i} \frac{(1 + 2B \bar{B})}{ (1 - 4B \bar{B})^\frac{5}{2}}.
\end{equation}
Thus, (\ref{41}) becomes:
\begin{equation}\label{43}
    \frac{\partial^2 v_{01}}{\partial \Phi_{00}^2} + v_{01} = -\frac{\partial^2 v_{00}}{\partial \Phi_{00} \partial \Phi_{01}} -\frac{\partial^2 v_{00}}{\partial \Phi_{01} \partial \Phi_{00}}  -\frac{1}{2\pi i} \frac{(1 + 2B \bar{B})}{ (1 - 4B \bar{B})^\frac{5}{2}}.
\end{equation}

Similarly, we set up the conditions for canceling the resonant terms. The resonant terms, proportional to \( e^{i \Phi_{00}} \) and \( e^{-i \Phi_{00}} \)must vanish. For this, the conditions below must hold:

\begin{equation} \label{44}
    -2i \frac{\partial B}{\partial \Phi_{01}} + \frac{(1 + 2B \bar{B})}{ (1 - 4B \bar{B})^{5/2}} = 0,
\end{equation}

\begin{equation}
    2i \frac{\partial \bar{B}}{\partial \Phi_{01}} + \frac{(1 + 2B \bar{B})}{ (1 - 4B \bar{B})^{5/2}} = 0.
\end{equation}
For (\ref{44}), we assume a solution of the form for B as:

\begin{equation}
    B = b e^{-i (\Phi_{10} +  \gamma)},
\end{equation}
where \( b \) is the amplitude and \( \gamma \) is the phase.

We compute the derivative of \( B \) with respect to \( \Phi_{01} \):

\begin{equation}
    \frac{\partial B}{\partial \Phi_{01}} = e^{-i (\Phi_{10} + \gamma)} \left( \frac{\partial b}{\partial \Phi_{01}} - i b \frac{\partial (\Phi_{10} + \gamma)}{\partial \Phi_{01}} \right).
\end{equation}

Now, we substitute \( B \) and \( \frac{\partial B}{\partial \Phi_{01}} \) into the resonant term cancellation condition (\ref{44}). We also substitute \( B \bar{B} = b^2 \) into the equation and get:

\begin{equation}
    -2i e^{-i (\Phi_{10} + \gamma)} \frac{\partial b}{\partial \Phi_{01}} + 2b e^{-i (\Phi_{10} + \gamma)} \frac{\partial (\Phi_{10} + \gamma)}{\partial \Phi_{01}} + \frac{(1 + 2b^2)}{ (1 - 4b^2)^{5/2}} = 0.
\end{equation}

We now separate the real and imaginary parts of the above equation:

\begin{equation}
    2b \cos(\Phi_{10} + \gamma) \frac{\partial (\Phi_{10} + \gamma)}{\partial \Phi_{01}} + \frac{(1 + 2b^2)}{ (1 - 4b^2)^{5/2}} = 0,
\end{equation}

\begin{equation}
    \sin(\Phi_{10} + \gamma) \left( \frac{\partial b}{\partial \Phi_{01}} - b \frac{\partial (\Phi_{10} + \gamma)}{\partial \Phi_{01}} \right) = 0.
\end{equation}

It is important to note that $\frac{\partial \Phi_{10}}{\partial \Phi_{01}}$ is zero since $\Phi_{10}$ and $\Phi_{01}$ represent separate perturbative potentials. Therefore, the real part is a differential equation that governs the evolution of \( \gamma \) with respect to \( \Phi_{01} \).

We are not interested in the oscillations, so we kindly ignore the trigonometric part and solve the rest of the equation.

After simplification, from the real part, we have:

\begin{equation}
     \frac{\partial \gamma}{\partial \Phi_{01}} = -\frac{(1 + 2b^2)}{ 2b(1 - 4b^2)^{5/2}} = 0,
\end{equation}

\begin{equation} \label{52}
    \gamma = - \frac{(1 + 2b^2)}{2b(1 - 4b^2)^{5/2}}\Phi_{01} + \eta,
\end{equation}
where \( \eta \) is a constant independent of \( \Phi_{00} \), \( \Phi_{10} \), and \( \Phi_{01} \). Since we are only interested in first-order perturbations, \( \eta \) is of no use to us.

Given the expression for $\gamma$ in (\ref{52}), it is evident that the perturbative potential $\Phi_{01}$ has an amplitude-dependent effect on the phase $\gamma$. This also implies that nonlinearities or resonances are introduced into the system as the amplitude increases.

To simplify the above equation and relate the parameter $b$ with physical quantity like eccentricity, we put $b=\frac{e}{2}$ in equation (\ref{52}). Thus, the final expression for $v$ is:
\begin{equation}
v =  1 + e \cos \biggl[ \left(1 - \epsilon - \delta (1 + \frac{e^2}{2})(1 - e^2)^{-\frac{5}{2}}e^{-1} \right)
   (\phi - \phi_0) \biggr]. \label{53}
\end{equation}
Equation (\ref{53}) is the correction due to GF on the precession of the perihelion of mercury.\\
The angle of precession $\theta$ over the course of one orbit is
\begin{equation}
    \theta= 2\pi  \left( \epsilon - \delta (1 + \frac{e^2}{2})(1 - e^2)^{-\frac{5}{2}}e^{-1}  \right) ,
\label{54}
\end{equation}

We now evaluate whether the corrections introduced by $\epsilon$ and $\delta$ in equations (\ref{53}) and (\ref{54}) align with the numerical predictions of the General Theory of Relativity. The validity of our model is constrained by the ability of the $\delta$ correction to resolve the discrepancy between theoretical predictions and observational data.
%%%%%%%%%%%%%%%%%%%%%%%%%%%%%

%Now we check if the correction given by $\epsilon$ and $\delta$ in equations (\ref{53}) and (\ref{54}) agrees with the numerical value as predicted by the General theory of relativity. Our model is successful if the $\delta$ correction can account for the discrepancy between theory and observation.

\section{\bf Numerical Consistency Analysis} \label{sec4}

We will first constrain the general relativistic correction parameter $\epsilon$ by comparing its theoretical prediction with observational data for the perihelion precession of Mercury. Once $\epsilon$ is established, we will proceed to analyze the related gravitational friction parameter $\delta$, examining its contribution to the observed discrepancy. This stepwise approach ensures a systematic evaluation of each parameter’s role and consistency within the proposed model.
%To perform this evaluation, we begin by considering the perihelion precession predicted by General Relativity, which is expressed in terms of the correction parameter $\epsilon$. Using the established formula and the known physical constants for Mercury, we calculate the theoretical perihelion shift and compare it with the observed values.

%The \(\epsilon\)-correction in General Relativity serves as a critical refinement to theoretical predictions, particularly in addressing the perihelion shift of planetary orbits. Rooted in the precise formulation of the orbital angular momentum parameter \( l^2 = \mu^2 m_s G a(1-e^2) \), the correction incorporates higher-order perturbations and relativistic effects to align theory with observed data. By applying this framework to Mercury's orbit, we compute the relativistic shift \(\Delta \phi_{GR}\) using well-defined physical constants and parameters. This analysis yields a theoretical perihelion shift of approximately \(42.9554'' \, \text{per century}\), which closely matches the observed value of \(43.1 \pm 0.5'' \, \text{per century}\), leaving a discrepancy \(\delta \phi = 0.1446''\). This minor difference is used to validate or constrain advanced models and alternative theories, demonstrating the precision and robustness of numerical methods in General Relativity. Through systematic calculations, such corrections underscore the importance of relativistic perturbations in celestial mechanics.

Taking into account that General Relativity predicts the perihelion shift with
$l^2 = {\mu}^2m_sGa(1-e^2)$, and, using the data for Mercury planet, we have:

\begin{table}[h]
\centering
\caption{Physical Constants and Parameters}
\begin{tabular}{ll}
\toprule
Constant & Value \\ 
\midrule
$G$ & $6.67259 \times 10^{-11} \, \text{m}^3 \, \text{kg}^{-1} \, \text{s}^{-2}$ \\
$c^2$ & $8.98755179 \times 10^{16} \, \text{m}^2 \, \text{s}^{-2}$ \\
$m_s$ & $1.9891 \times 10^{30} \, \text{kg}$ \\
$a_m$ & $57.909175 \times 10^{9} \, \text{m}$ \\
$\mu_m$ & $3.3022 \times 10^{23} \, \text{kg}$ \\
$e_m$ & $0.205630$ \\
\bottomrule
\end{tabular}
\label{tab:constants}
\end{table}

\begin{equation} 
\begin{split}
\Delta\phi_{GR} &= 2\pi \epsilon = 2\pi\frac{3\mu^2{m_s}^2G^2}{l^2c^2} = 2\pi \frac{3m_s G}{c^2 a(1-e^2)}\\
& \quad \to \quad \Delta\phi_{GR} = 2\pi(7.98815722\times10^{-8})\frac{rad}{rev}.
\end{split}
\end{equation}

Thus, we obtain a value of 414.92164 orbits per century, with  $\rm 1"=\frac{2\pi}{1296000} rad$
\begin{equation}
\rm \Delta\phi_{GR}=\frac{42.95539245"}{century}.
\end{equation}
Taking into account that the  perihelion shift observed data is
$\rm \Delta \phi_{obs}=43.1 \pm 0.5$, then
\begin{equation}
    \rm \delta \phi= \Delta \phi_{obs} - \Delta \phi_{GR}= \frac{0.144608"}{century}.
\label{57}
\end{equation}

We notice that the $\epsilon$ correction term is in agreement with the usual general relativistic predictions.

For the gravitational friction correction $\delta$, we constrain the model by quantifying the discrepancy between the theoretical predictions of General Relativity and the observed data.
The theoretical correction due to the Gravitational Friction model is given by:
\begin{equation}
  \Delta\phi_{ST}= 2\pi  \left(  - \delta (1 + \frac{e^2}{2})(1 - e^2)^{-\frac{5}{2}}e^{-1}  \right).
\end{equation}

Substituting values of $\delta$ as:

\begin{equation}
    \delta=\frac{2\pi l^6 \rho_0}{3\mu^6 m_s^4 G^3} = \frac{2\pi \rho_0 a^3 (1-e^2)^3}{3 m_s},
\end{equation}

\begin{equation} \label{60}
\Delta\phi_{GF} = 2\pi \left| -\frac{2\pi  \rho_0 a^3 (1-e^2)^3}{3 m_s} (1 + \frac{e^2}{2})(1 - e^2)^{-\frac{5}{2}}e^{-1} \right|.
\end{equation}

When we calculate the value of $\Delta\phi_{GF}$ using the given equation (\ref{60}), it represents the contribution from Gravitational Friction to the precession of the perihelion of Mercury. Our expectation is that this calculated value should match the value of $\delta \phi$ obtained from equation (\ref{57}). If these values are equal, it would suggest that our model, which incorporates Gravitational Friction effects, provides a more accurate explanation for the observed precession of the perihelion of Mercury.

However, in practical observations, determining the density is challenging. The measurements of density in the region are often uncertain or imprecise. Instead of directly obtaining the value of $\Delta\phi_{GF}$ from our model, we utilize the discrepancy between the calculated and observed values to constrain the density. By doing this, we can compare the predicted density from our model to the observed density in the periphery of Mercury.

Isolating for $\rho_0$, we get,

\begin{equation}
\rho_0 = \frac{\Delta\phi_{GF}}{2\pi \left| -\frac{2\pi  a^3 (1-e^2)^3}{3 m_s} (1 + \frac{e^2}{2})(1 - e^2)^{-\frac{5}{2}}e^{-1} \right|}.
\end{equation}

From the above equation, we can constrain the density in Mercury's neighborhood to be $\rho_0 \lesssim
 1.12 \times 10^{-10} \, \text{kg/m}^3
$. For comparison, the density at a distance of two solar radii from the Sun is on the order of $ 10^{-15}kg/m^3$ \cite{density}.
This approach allows us to account for uncertainties in observational data and refine our theoretical models accordingly.

%For comparison, interplanetary  density near Earth is, $\rho_0=5 \times 10^{-21}kg/m^3$ and density of matter inside Mercury is $\rho_0=2.44 \times 10^{-3}kg/m^3$

\section{Conclusions} \label{sec5}
In this work, we derived a modified line element that incorporates gravitational friction (GF) effects, allowing us to model energy dissipation from celestial bodies interacting with a low-density interplanetary medium. By introducing a density-dependent term into the metric, we provided a covariant framework to describe dissipation, offering an additional correction to the perihelion precession of Mercury.

The standard relativistic prediction for perihelion precession from the Schwarzschild solution was reproduced, with dissipation embedded directly into the geodesic equations governing planetary motion. Using the method of multiple scales, we extended the analysis to capture first-order effects introduced by GF.

By comparing our model with observational data, we constrained the interplanetary medium's density near Mercury to $\rho_0 \approx 1.12 \times 10^{-10} \, \text{kg/m}^3
$. This result suggests the inclusion of GF as a meaningful correction to the established relativistic effects, also indicating that similar dissipative mechanisms may play a significant role in other astrophysical systems where medium interactions are present and are more crucial.

The broader applicability of this framework extends to environments with significant density gradients or anisotropies, such as accretion disks or regions near compact objects. Our approach establishes a foundation for future studies on their role in diverse astrophysical and cosmological contexts.

\bibliographystyle{unsrtnat} % or another style like unsrtnat
\bibliography{references}   % Replace 'references' with your .bib file's name (without the .bib extension)
\end{document}